\documentclass{template/svproc}

\usepackage{tikz}
\usepackage{amsmath}
\usepackage{tcolorbox}
\tcbuselibrary{breakable, skins}
\usepackage{caption}
\usepackage{paralist}
\usepackage{url}
\usepackage{booktabs}
\usepackage[hidelinks]{hyperref}
\usepackage{longtable}
\usepackage{array}

\begin{document}
\mainmatter              
\title{Actionable Cybersecurity Notifications for Smart Homes: A User Study on the Role of Length and Complexity}

\titlerunning{Actionable Cybersecurity Notifications for Smart Homes}  
%
\author{Victor J{\"u}ttner\inst{1,2} \and Charlotte S. L{\"o}ffler\inst{3} \and Erik Buchmann\inst{1,2}}

\authorrunning{J{\"u}ttner et al.} 

\tocauthor{Victor J{\"u}ttner, Charlotte S. L{\"o}ffler, Erik Buchmann}

\institute{
Dept. of Computer Science, Leipzig University, Germany\\
\email{\{juettner, buchmann\}@informatik.uni-leipzig.de}
\and
Center for Scalable Data Analytics and Artificial Intelligence (ScaDS.AI) Dresden/Leipzig, Germany
\and
Cognitive Psychology, University of Kassel, Kassel, Germany\\
\email{charlotte-loeffler@uni-kassel.de}
}

\maketitle              

\begin{abstract}

The proliferation of smart home devices has increased convenience but also introduced cybersecurity risks for everyday users, as many devices lack robust security features. Intrusion Detection Systems  are a prominent approach to detecting cybersecurity threats. However, their alerts often use technical terms and require users to interpret them correctly, which is challenging for a typical smart home user. Large Language Models can bridge this gap by translating IDS alerts into actionable security notifications. However, it has not yet been clear what an actionable cybersecurity notification should look like. In this paper, we conduct an experimental online user study with 130 participants to examine how the length and complexity of LLM-generated notifications affect user likability, understandability, and motivation to act. Our results show that intermediate-complexity notifications are the most effective across all user groups, regardless of their technological proficiency. Across the board, users rated beginner-level messages as more effective when they were longer, while expert-level messages were rated marginally more effective when they were shorter. These findings provide insights for designing security notifications that are both actionable and broadly accessible to smart home users.


\end{abstract}

%

\section{Introduction}

\label{sec:intro}

Automating everyday life with smart-home appliances is convenient, but it also introduces a plethora of cybersecurity risks~\cite{BUILGIL2023107770}. A state-of-the-art approach for detecting such threats is to deploy an intrusion detection system (IDS)\cite{AnthiIDS2018}, which monitors network traffic and generates alerts when suspicious activity occurs. However, interpreting these alerts typically requires cybersecurity expertise -- an expertise that the average smart-home user does not possess. For example, consider alerts such as \textit{"INDICATOR-SCAN ipEye SYN scan"} (Snort community rules\cite{snort_ruleset}), \textit{"HTTP Response abnormal chunked for transfer encoding"} (Suricata rules~\cite{suricata_ruleset}), or \textit{"ET CNC Feodo Tracker Reported CnC Server TCP group 1"} (Emerging Threats rules~\cite{et_ruleset}). Responding to these alerts requires placing them in the context of the current network environment, filtering out false positives, and implementing appropriate countermeasures. In a smart-home setting, there is the added challenge that users must be willing to engage with such alerts during their leisure time.

This presents a fundamental limitation to the practical use of IDS solutions in private home networks. Cybersecurity threats, device types, and appropriate responses evolve rapidly, making it difficult to rely on static, human-authored explanations embedded in IDS rule sets.
Recent work has explored the use of large language models (LLMs) to bridge the gap between technical IDS alerts and user-friendly communication. LLMs can generate adaptable, context-aware, and actionable notifications from raw IDS outputs~\cite{juettner2024chatids,hoffmann2024ChatSEC}. Here, “actionable” refers to notifications that help users distinguish real threats from false positives, provide enough context to understand the situation, and guide appropriate responses. In this way, LLMs have the potential to empower smart-home users to secure their networks—even without technical backgrounds.

While prior work has addressed how to factor in user motivation and skill levels~\cite{Boss2015}, there is still a gap in understanding what actionable notifications should look like for non-expert users in private settings. In particular, there is limited understanding of how length and complexity of LLM-generated notifications impact user engagement. 
Thus, our research question is:

\textbf{How should LLM-generated security notifications for smart homes be designed in terms of length and complexity to effectively engage users with varying technical proficiencies?}

We approach this question with a quantitative approach to capture broad patterns in user preferences across diverse levels of technical proficiency. This allows statistically grounded recommendations for user-centered notification design. 
In particular, we conducted an experimental online user study with 130 participants recruited through Prolific. Our goal was to derive generalizable design principles for actionable security notifications by systematically varying their \textit{length} (short, long) and \textit{complexity} (beginner, intermediate, expert) across a set of 60 LLM-generated examples. Each participant rated a stratified subset of 30 notifications on \textit{likability}, \textit{understandability}, and \textit{motivation to act}. The notifications were generated using OpenAI’s GPT-4o~\cite{openai2024gpt4o} and the Snort IDS Community Ruleset~\cite{snort_ruleset}, with a tailored prompt template. 
The study was implemented using jsPsych~\cite{deLeeuw2023} and hosted on a JATOS server~\cite{lange2015jatos} provided by MindProbe~\cite{mindprobe}.

To the best of our knowledge, this is the first study to apply psychological methods to evaluate how 
users perceive LLM-generated actionable cybersecurity notifications. We found that notifications with intermediate complexity were consistently rated highest in likability, understandability, and motivation to act, across all levels of user expertise. The effect of length was more nuanced: longer notifications were more effective when written at the beginner level, while shorter versions of expert-level notifications were marginally preferred. These findings offer concrete guidelines for designing effective, user-centered security notifications, and can inform future work on human-centered cybersecurity beyond LLM-generated content.

Section~\ref{sec:related} reviews related work, Section~\ref{sec:method} details the creation of security notifications, Section~\ref{sec:study} outlines the experimental methodology, Section~\ref{sec:result} presents the results, Section~\ref{sec:discussion} discusses the findings, and Section~\ref{sec:conclusion} concludes.

\section{Related Work}
\label{sec:related}

This section summarizes relevant aspects of smart-home cybersecurity and cybersecurity notifications, and identifies our research gap.

\subsection{Smart Homes and Cybersecurity}
Smart homes introduce significant cybersecurity challenges due to the widespread integration of Internet of Things (IoT) devices. Many devices lack robust security features, leaving them susceptible to attacks and exposing sensitive user data. The key challenges can be summarized as follows:

\begin{itemize}
\item[\textbf{Device Vulnerabilities}]
Many smart home devices operate with limited computational resources and are often designed without security as a primary consideration. These devices frequently lack regular firmware updates, which makes them prime targets for attackers~\cite{smart-device-vulns}.

\item[\textbf{Network Security}]
The interconnected nature of smart homes creates a complex attack surface. Compromising a single device can grant attackers access to the entire home network, amplifying the risks~\cite{smart-network-issues}.

\item[\textbf{User Awareness}]
Users often lack the technical knowledge required to implement or maintain cybersecurity measures. Studies highlight that many users remain unaware of potential threats, further exacerbating the risks~\cite{Pattnaik2022ASO}.
\end{itemize}




\paragraph{Intrusion Detection} Considering these challenges, IDS are an important security measure. In this work, we focus on signature-based IDS, which compare network traffic against a database of attack signatures, providing a first line of defense against potential intrusions. However, traditional IDS solutions produce highly technical outputs that are difficult for average users to interpret, underscoring the need for more user-centered approaches~\cite{nist-ids}.

\subsection{Security Notifications}
Effective security notifications play a critical role in mitigating cybersecurity threats by translating technical information into actionable insights for users. The design of these notifications must account for human factors, including cognitive and emotional responses.

\textit{Human-Centered Security Design}
Cranor emphasizes the importance of "humans in the loop" in security systems, advocating for designs that support user engagement and reduce the likelihood of errors~\cite{cranor-human-in-loop}. Similarly, Zimmermann et al. propose a shift from the "human-as-problem" mindset to a "human-as-solution" perspective, recognizing the potential of users to actively contribute to cybersecurity when supported by intuitive systems~\cite{ZIMMERMANN2019169}.

\textit{Behavioral Frameworks}
Protection Motivation Theory (PMT) highlights the importance of combining fear appeals with actionable solutions (response efficacy) and user confidence (self-efficacy)~\cite{rogers1983cognitive}. The Communication-Human Information Processing (C-HIP) model by Wogalter further emphasizes the need for notifications to capture attention, enhance comprehension, and motivate action~\cite{wogalter2018communication}. These frameworks underline the necessity of designing notifications that align with user mental models and decision-making processes.

\textit{Challenges in Notification Design}
Despite the theoretical frameworks, practical challenges persist. Users often encounter notifications that are overly technical, leading to confusion and disengagement~\cite{warning-study}. Habituation to frequent alerts and false positives further erode trust in notifications. To rebuild user confidence, notifications must be accurate and contextually relevant~\cite{scaring-warning}.

\textit{Emotional and Cognitive Factors}
The emotional impact of notifications is also critical. Conrad et al. demonstrate that fear-based messages can motivate protective actions but must be carefully balanced to avoid overwhelming users~\cite{warning-eeg}. Zimmermann et al. further explore the interplay of emotional, cognitive, and social factors in cybersecurity, emphasizing the importance of addressing user emotions to improve engagement~\cite{zimmermann2024beyondFear}.

\subsection{LLMs in Usability and Security}
LLMs have demonstrated their potential in enhancing user-centered communication, including in cybersecurity contexts.

\textit{Applications in Cybersecurity}
LLMs have been employed for tasks such as code generation and debugging, as seen in tools like Microsoft Copilot~\cite{microsoft2024copilot}. Research has also shown that LLMs can enhance code security and data privacy, outperforming traditional methods in many cases~\cite{YAO2024100211}.

\textit{Challenges and Limitations}
Despite their promise, LLMs face challenges such as hallucinations, where the model generates inaccurate or nonsensical information~\cite{stochastic-parrots}. In critical applications like cybersecurity, these inaccuracies can undermine user trust and lead to ineffective responses~\cite{Hicks2024}. This highlights the need for careful evaluation and design when integrating LLMs into user-facing systems.

\subsection{Smart Home Security Notifications}
Existing academic literature on smart home security notifications reveals a spectrum of approaches, ranging from systems that actively notify users to those that focus solely on intrusion detection without explicit user engagement.

\paragraph{Systems with Notifications}
Several systems incorporate user notifications as part of their security framework: Aegis+: A smartphone application delivers customizable security notifications detailing malicious events, affected devices, and their locations~\cite{aegis}. Dynamic Risk Assessment Framework (DRAF): This system automates the identification of IoT attacks and provides notifications adjustable to users' preferred risk levels~\cite{draf}. ChatIDS: An IDS-based system that integrates LLMs to create user-friendly notifications, translating technical alerts into actionable insights~\cite{juettner2024chatids}. PITI: A hybrid IDS implementation that notifies users through sound and text alerts, providing details such as attack type and device IPs~\cite{piti}. CogSec: A repository of over 100 cyberattacks with descriptions designed to make security threats understandable for non-technical users~\cite{user-centric-threat-model}.

\paragraph{Systems without Notifications}
Some frameworks focus on enhancing smart home cybersecurity without explicitly addressing user notifications: GHOST: A multi-layered cybersecurity framework providing real-time risk control~\cite{ghost,internet-of-threats-ghost}. In-Hub Security Manager: A centralized system managing device status, software updates, and traffic filtering~\cite{hub2027}. IOTFLA: A federated learning-based architecture enhancing privacy and security without engaging users directly~\cite{iotfla}.

\subsection{Research Gap and Contribution}
While prior research has explored design principles for security notifications and the implementation of notification systems in smart homes, there is a lack of actionable advice, personalization, and clear guidelines on how to tailor security notifications for diverse users. 
To address this gap, we conduct an experimental user study evaluating how LLM-generated security notifications derived from IDS alerts—with varying length and complexity—impact likability, understandability, and motivation to act. 
Our findings provide empirical insights and actionable guidelines for designing notifications tailored to users with different technical proficiencies in smart home environments.

\section{Building the Security Notifications}
\label{sec:method}


This section explains how we designed the security notifications for our study.

\subsection{Security Notification Design} 
We design security notifications with varying \textbf{length} and \textbf{complexity} to serve as the basis of our experimental user study. We systematically manipulate these two independent variables and generate notifications accordingly.

Our security notifications follow a structured format consisting of a greeting, an introduction, instructional steps, and a closing. The introduction explicitly mentions the specific smart home device. To manipulate the length, we vary the number of instructional steps—either two or four—depending on whether the message is short or long.

To manipulate complexity, we adjust the content of the instructional steps. Complexity is defined in terms of the technicality of the language and the depth of knowledge required to follow the instructions. Specifically, complexity increases with the use of technical terms and more advanced actions that require greater expertise.
We define three different user types, each of which corresponds to varying levels of complexity:
\begin{itemize}
\item[\textbf{Beginner:}] Can perform basic tasks such as turning devices on and off, seeking help, and following simple instructions with minimal technical knowledge. For beginners, the instructional steps are straightforward, with minimal technical jargon and simple actions.
\item[\textbf{Intermediate:}] Can manage tasks such as device and router resets, installing updates, and changing passwords. For intermediate users, the instructional steps are more detailed, involving some technical terms and requiring a moderate level of prior knowledge.
\item[\textbf{Expert:}] Possesses IT-expert knowledge and only needs assistance with cybersecurity-specific issues. For experts, the instructions are highly technical, containing specialized terms and requiring more prior knowledge. The steps contain more complex actions that experts are expected to understand and execute.
\end{itemize}

The greeting and introduction are approximately 40 words long, depending on the smart device. Each instructional step is designed to be around 30 words. 
We manipulate the overall length of the notifications as follows:

\begin{itemize}
    \item[\textbf{Short:}] Two steps, approximately 100 words in total.
    \item[\textbf{Long:}] Four steps, approximately 160 words in total.
\end{itemize}

Figure~\ref{fig:example-sec-notification} shows an exemplary short intermediate security notification. Further examples of security notifications for short beginner, short expert, and long intermediate notifications can be found in Appendix~\ref{sec:sec-notifications}. All examples are based on the "Amazon Echo Dot 5th Gen device" for comparison purposes.

\begin{center}
\begin{tcolorbox}[breakable, enhanced, sharp corners, colback=white, colframe=black, boxrule=0.5pt, left=1mm, right=1mm, top=1mm, bottom=1mm]
\small
Hello,\\
Your smart home security system has found a possible security problem with your Amazon Echo Dot 5th Gen, a smart speaker. Please follow these suggestions to fix the issue and secure your device and network:

\begin{itemize}
    \item\textbf{Power Cycle and Update}: Unplug your Amazon Echo Dot for 30 seconds. Plug it back in, then use the Alexa app to check for and install any available firmware updates to ensure it is up-to-date.
    \item \textbf{Change Wi-Fi Password and Reconnect}: Access your router settings to change your Wi-Fi network password. Afterward, reconnect your Echo Dot to the newly secured network to mitigate unauthorized access.
\end{itemize}

Best regards,\\
Your Smart Home Security Team
\end{tcolorbox}
\captionof{figure}{Short Intermediate Security Notification}
\label{fig:example-sec-notification}
\end{center}

\subsection{Development of Prompt Templates}
We developed our prompt templates iteratively using the OpenAI Web interface for GPT-4o. We refined the basic prompt template over 20 iterations. Afterward, we conducted approximately 5 additional iterations per template to adjust for varying levels of complexity. After finalizing the templates, we transitioned to working with the OpenAI API.

To tailor our security notifications to different user types and lengths, we developed six distinct prompt templates. Each user type—beginner, intermediate, and expert—has a template for both short and long messages. We employed several prompting techniques to create our templates:

\begin{itemize}
    \item[\textbf{Scenario Description}:] Scenario with a person, environment, and incident.
    \item[\textbf{Role Assignment}:] The LLM is assigned the role of a cybersecurity expert.
    \item[\textbf{Task Description}:] Explaining the incident and providing assistance.
    \item[\textbf{User Type Description}:] We detail the user's technical knowledge.
    \item[\textbf{Format Specifications}:] The number of instructional steps and word count.
    \item[\textbf{Example Notification}:] Sample security notification to serve as a template.
\end{itemize}

The descriptions of the user type and format specifications vary between templates to accommodate the different knowledge levels and the required length of the message. Before sending the prompt template to the LLM, we replace the placeholders for device and IDS alert. The prompt templates for the beginner-short, intermediate-short, and exper-short security notifications are in Appendix~\ref{sec:prompt}. The templates for the long security notifications only differ by stating that four instead of two instructional steps are needed.


\subsection{Generation of Security Notifications}
We let an LLM generate security notifications from Snort rules, based on a diverse set of smart home devices, and inspect the quality of the notifications.

\paragraph{LLM Setup}
For generating our security notifications, we utilize OpenAI's GPT-4o-2024-05-13 model through the OpenAI API~\cite{openai2024gpt4o}. We chose GPT-4o for a few reasons: it has proven performance in understanding and generating natural language; using a single model keeps our experiments consistent; and the accessibility of OpenAI's API makes generating several security notifications easier.

We use OpenAI's default settings:
both Top P and Temperature are set to 1, while Frequency and Presence penalties are set to 0. These settings ensure a balance between creativity and coherence in the generated text, without imposing additional constraints on the model's output. Additionally, we use only the user prompt without incorporating a system prompt, aligning our setup with the default settings available on the OpenAI web interface for the GPT-4o model.

\paragraph{Selection of Snort Rules}
To simulate attacks for our security notifications, we select ten rules from the Snort Community Ruleset~\cite{snort_ruleset}. Each rule corresponds to a specific type of attack that could threaten a smart home network. We categorize the rules into different classtypes~\cite{snort_classtypes}, assigning each a severity rating from 1 to 4, where 1 denotes the highest severity and 4 the lowest. We have chosen rules that vary in severity and attacktype to represent a wide range of threats.

Although Snort rules contain extensive information, we focus solely on the alert message of each rule for our purpose. This alert message provides a concise description of the detected attack, such as ``\texttt{PROTOCOL-ICMP Nemesis v1.1 Echo}.'' Appendix~\ref{sec:alert-messages} presents a detailed overview of the selected rules, including the Snort ID (SID), the alert message used to generate the warning, an explanation of the attack, and its severity.

\paragraph{Selection of Smart Home Devices}
We selected ten different smart home devices, varying in brand, functionality, and sensor type. These devices include popular brands to ensure relevance. The chosen devices feature a range of sensors such as microphones, cameras, and thermometer, as well as devices like locks, door sensors, and lamps. Encompassing both comfort and security functionalities. Appendix~\ref{sec:devices} provides an overview of all selected devices along with descriptions.

The type of device could influence how the LLM phrases the security notifications. For instance, security devices like smart locks might prompt more urgent and severe warnings compared to devices like smart light bulbs. To control for this potential bias, we  selected a diverse set of devices to suppress this effect.

\paragraph{Compilation of Security Notification Dataset}
For our dataset, we created 10 security notifications for each of the 6 user-length categories, resulting in a total of 60 security notifications. Each notification incorporated one of the ten selected devices and one of the ten Snort rules. To ensure diversity, we varied the device-rule combinations across the categories, covering a wide range of incidents.

\begin{figure}[h]
    \centering
    \includegraphics[width=.7\textwidth]{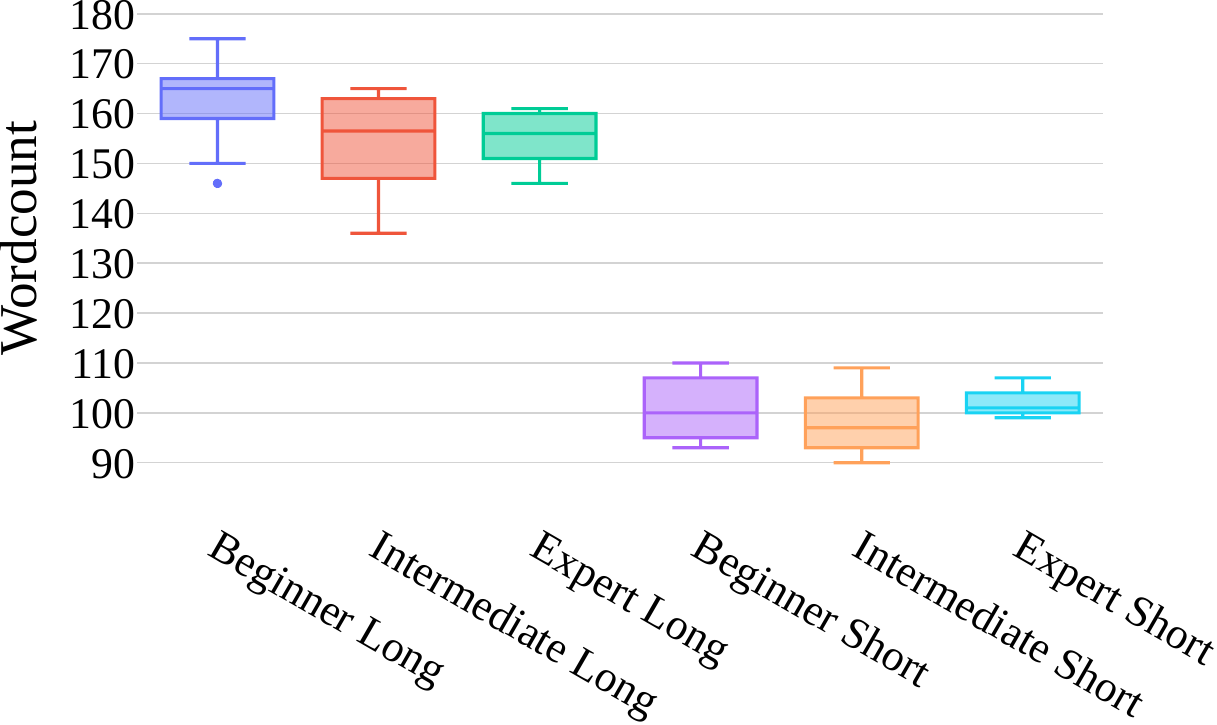}
    \caption{Wordcounts of all Security Notifications for each User-Length-Category}
    \label{fig:wordcount}
\end{figure}

To maintain length accuracy, we used a word counter to verify that each message met the specified length criteria for its category, as shown in Figure~\ref{fig:wordcount}. Finally, we conducted a qualitative check by manually reviewing each security notification to ensure the instructions fit the defined user types.

\section{User Study}
\label{sec:study}

We developed our experiment using jsPsych~\cite{deLeeuw2023} and hosted it on a JATOS server~\cite{lange2015jatos} provided by MindProbe~\cite{mindprobe}. This section provides an overview of participant demographics and the experimental design.

\subsection{Participants}

We recruited participants for this experimental user study through Prolific. A total of $N= 130$ participants completed the study, with each receiving a compensation of £2.50 for their participation. A pretrial with 10 participants on Prolific indicated that participants required an average of approximately 25 minutes to complete the survey.

\begin{figure}[h]
    \centering
    \includegraphics[width=.7\textwidth]{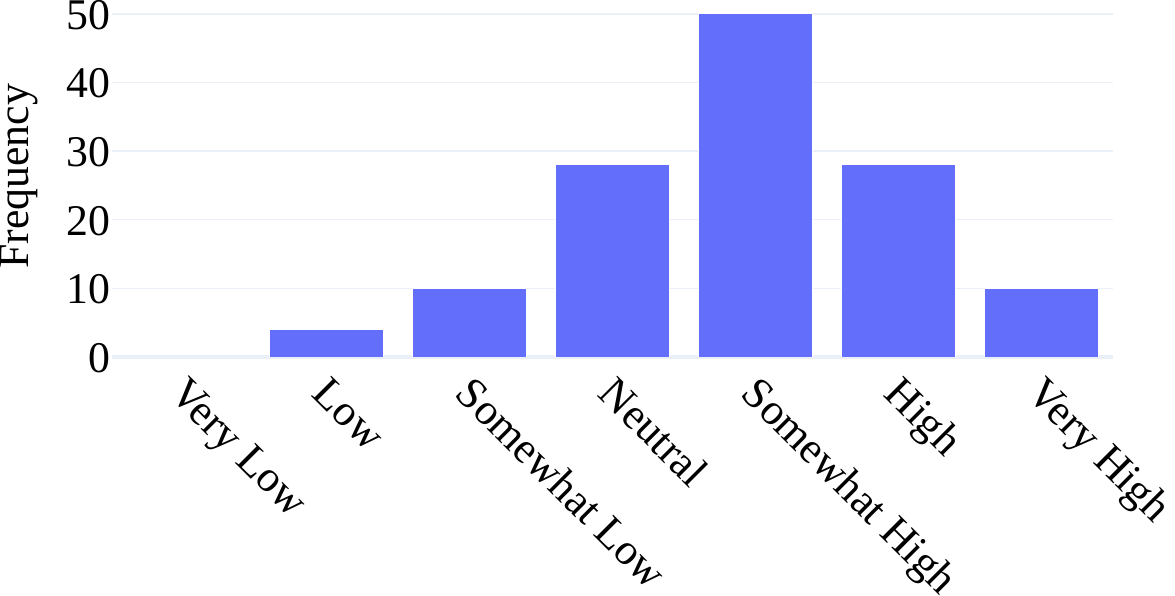}
    \caption{Participants self-rated technological proficiency.}
    \label{fig:tech_prof}
\end{figure}

We conducted a power analysis using G*Power~\cite{faul2009statistical} before data collection to determine the necessary sample size for detecting meaningful differences between conditions (notification length and user expertise level). Using a two-tailed test, an anticipated effect size of 0.25, a significance level of $\alpha = 0.05$, and a power of $0.80$, the analysis indicated that 128 participants were required. Based on these parameters, we deemed the final sample size of 130 participants sufficient for this experimental user study. \textbf{Demographic information} of the participants in the maintrial, selected through Prolific:

\begin{itemize}
    \item[\textbf{Age:}] 18–83 years (median: 42) Distribution: 18-28 years~(12\%), 28-38 years~(28\%), 38-48 years~(25\%), 48-58 years~(22\%), 53+ years~(13\%).
    \item[\textbf{Gender Identity:}] Male (62\%), Female (38\%)
    \item[\textbf{Technical Proficiency:}] Very Low (0\%), Low (3\%), Somewhat Low (8\%), Neutral (22\%), Somewhat High (38\%), High (22\%), Very High (8\%) (Figure \ref{fig:tech_prof}).
    \item[\textbf{Country of Residency:}] UK (85\%), US (12\%), Ireland (3\%)
\end{itemize}

 Since we did not target vulnerable groups and only individuals aged 18 or older were recruited through Prolific, formal Institutional Review Board approval was not required. This study adhered to university ethics guidelines and national data protection regulations. Participants first reviewed the study description on Prolific before proceeding to our experimental user study on MindProbe, hosted on an EU server to comply with data protection laws. Informed consent was obtained before the study began, and participants could withdraw at any time without consequences. 
 Data was anonymously collected with random identifiers.

\subsection{Experimental Design} 
Each participant viewed a total of 30 security notifications, which we selected using stratified random sampling from the dataset of 60 notifications. To ensure balanced exposure, each participant viewed five notifications from each category, with the selection and order of notifications randomized for each participant to prevent order effects.

After viewing each security notification, we asked participants 
questions corresponding to our dependent variables (Liking, Understandability, and Motivation to Act), rated on a 7-point Likert scale and presented as radio buttons.

\begin{itemize}
    \item[] \textbf{``How much do you like this security notification?''} \\(1 = not at all, 7 = very much)
    \item[] \textbf{``How understandable is this security notification?''} \\(1 = not at all, 7 = very much)
    \item[] \textbf{``How likely would you take action if you received this security notification?''} \\(1 = very unlikely, 7 = very likely)
\end{itemize}

We also asked for a fourth dependent variable concerning the trustworthiness of the notifications. However, we did not receive significant results. For the sake of brevity, we omitted this aspect in the paper. 

We displayed each notification for a minimum duration based on its length—20 seconds for short notifications and 30 seconds for long notifications. Participants could not proceed until the minimum viewing time had elapsed, ensuring they engaged with each notification adequately before moving on. We presented notifications in a visually distinct box with a light blue background and a black border to differentiate them from the rating scales below.

At the end of the experiment, we asked participants to provide demographic information, including gender (with options such as "male," "female," "non-binary," and "prefer not to say"), age, and self-assessed technical proficiency (on a seven-point scale from "very low" to "very high").

\section{Results}
\label{sec:result}

We analyzed the data using IBM SPSS Statistics (Version 29). Effect sizes for repeated measures analysis of variances (ANOVAs) were calculated using partial eta squared ($\eta^2$) and interpreted as small ($\eta^2 = 0.01$), medium ($\eta^2 = 0.06$), and large ($\eta^2 = 0.14$) based on Cohen’s (1988) guidelines~\cite{cohen1988}. For paired-samples t-tests, effect sizes were calculated using Cohen's $d_z$ with $d_z = 0.20$, $d_z = 0.50$, and $d_z = 0.80$ representing small, medium, and large effects, respectively.

\subsection{Likability}
We conducted a $3\times2$ within-subjects ANOVA to examine the effects of notification complexity (beginner vs. intermediate vs. expert) and notification length (short vs. long) on likability ratings. The analysis revealed a small to medium-sized main effect of notification complexity, with intermediate-level notifications ($M = 3.77$, $SE = 0.10$) being rated significantly more likeable than beginner ($M = 2.96$, $SE = 0.12$) and expert-level notifications ($M = 3.23$, $SE = 0.11$), $F(1, 129) = 52.79$, $p < .001$, partial $\eta^2 = 0.29$. 
This pattern was evident among participants independent of their levels of technical proficiency.

Further, a small but significant main effect of notification length was observed, indicating that longer notifications ($M = 3.39$, $SE = 0.11$) received higher likeability ratings than shorter notifications overall ($M = 3.25$, $SE = 0.10$), $F(1, 129) = 4.28$, $p = .041$, partial $\eta^2 = 0.03$.

\begin{figure}[h]
    \centering
    \includegraphics[width=.7\textwidth]{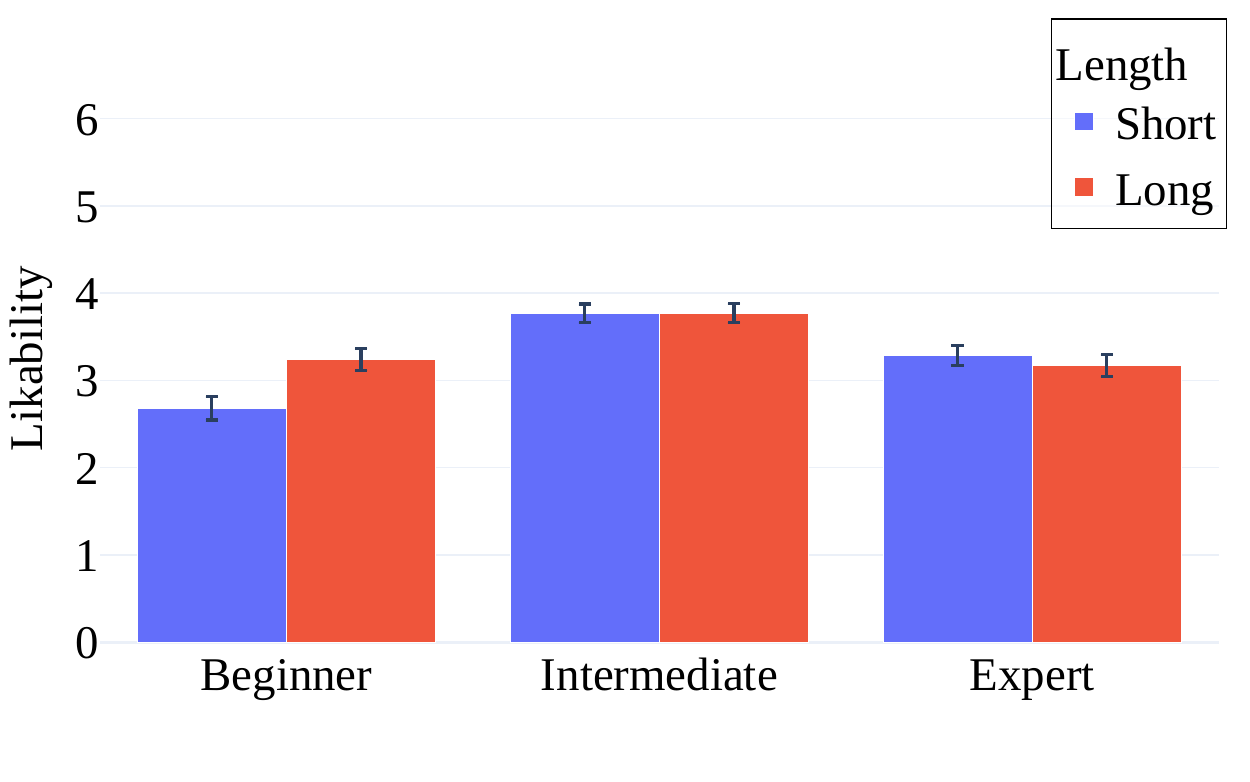}
    \caption{Likability ratings by notification complexity and length, showing intermediate notifications as most likeable. Error bars represent standard errors of the mean.}
    \label{fig:likability}
\end{figure}

A small interaction effect between notification complexity and length emerged, $F(1, 129) = 30.03$, $p < .001$, partial $\eta^2 = 0.19$. The interaction was constituted by the pattern that at the beginner level, longer notifications ($M = 3.24$, $SE = 0.13$) received higher likability ratings than shorter notifications $M = 2.68$, $SE = 0.14$), $t(129) = 5.86$, $p < .001$, $d_z = 0.52$. In contrast, at the expert level, shorter notifications ($M = 3.28$, $SE = 0.11$) were rated marginally higher in likeability than longer notifications ($M = 3.17$, $SE = 0.12$). However, the difference was not statistically significant, $t(129) = 1.34$, $p = .091$, $d_z = 0.12$.


In summary, our analysis found that participants liked intermediate-level notifications best, outperforming both beginner and expert-level notifications, regardless of technical proficiency. Additionally, participants generally preferred longer notifications, especially beginners. We observed a complex interaction between notification complexity and length (see Figure~\ref{fig:likability}): At the beginner level, longer notifications were rated more likeable, whereas at the expert level, shorter notifications were rated slightly higher, though this difference was not significant.

\subsection{Understandability}
We observed a similar pattern regarding the understandability of the notifications. A 3 (beginner vs. intermediate vs. expert) $\times$ 2 (short vs. long) ANOVA on the understandability ratings revealed a small main effect of notification complexity, with intermediate notifications  ($M = 4.52$, $SE = 0.09$) being rated as more understandable compared to beginner ($M = 4.14$, $SE = 0.10$) and expert-level notifications  ($M = 3.60$, $SE = 0.11$), $F(1, 129) = 76.62$, $p < .001$, partial $\eta^2 = 0.37$. Again, this pattern emerged consistently across participants with both high and low levels of technical proficiency.

Additionally, a small main effect of notification length was observed, indicating that shorter notifications ($M = 4.22$, $SE = 0.09$) were rated as more understandable than longer notifications overall ($M = 3.95$, $SE = 0.10$), $F(1, 129) = 24.37$, $p < .001$, partial $\eta^2 = 0.16$.

Further, there was a small interaction effect between complexity and length, $F(1, 129) = 13.37$, $p < .001$, partial $\eta^2 = 0.09$. Post-hoc analyses revealed that the understandability advantage of shorter notifications was most pronounced for expert-level notifications ($M_{\text{short}} = 3.82$, $SE_{\text{short}} = 0.11$; $M_{\text{long}} = 3.37$, $SE_{\text{long}} = 0.12$), $t(129) = 6.19$, $p < .001$, $d_z = 0.55$, while no such advantage was observed for beginner-level notifications ($M_{\text{short}} = 4.16$, $SE_{\text{short}} = 0.11$; $M_{\text{long}} = 4.12$, $SE_{\text{long}} = 0.11$), $t(129) = 0.50$, $p = .622$, $d_z = 0.04$.


\begin{figure}[h]
    \centering
    \includegraphics[width=.7\textwidth]{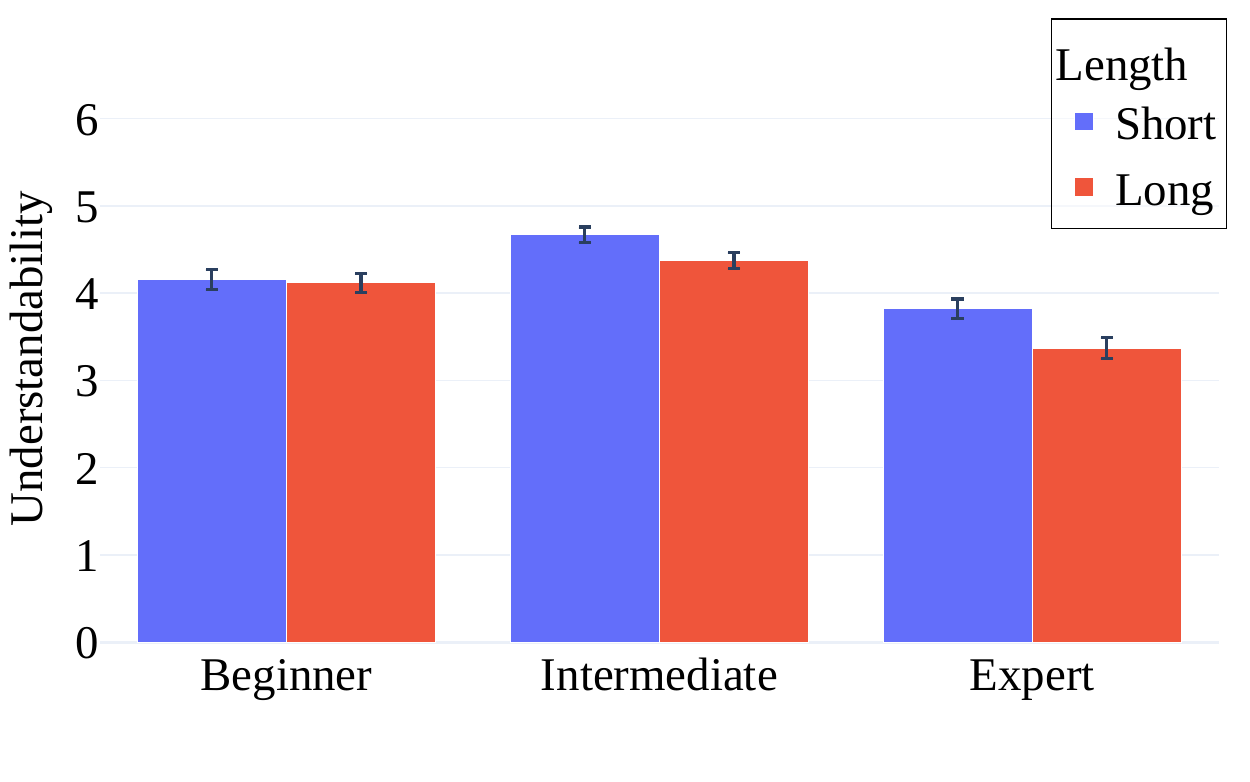}
    \caption{Understandability ratings by notification complexity and length, showing intermediate notifications as most understandable. Error bars represent standard errors of the mean.}
    \label{fig:understand}
\end{figure}

In summary, our analysis revealed that participants rated intermediate-level notifications as the most understandable, followed by beginner and expert-level notifications. Participants generally considered shorter notifications more understandable, with this effect being especially noticeable for expert-level notifications (see Figure~\ref{fig:understand}). Participants found no difference in understandability between short and long notifications at the beginner level.

\subsection{Motivation to Act}
Finally, we performed a 3 (beginner vs. intermediate vs. expert) $\times$ 2 (short vs. long) ANOVA on the aggregated motivation-to-act ratings. Again, a small main effect of notification complexity emerged, with intermediate-level notifications ($M = 4.34$, $SE = 0.09$) eliciting higher motivation to act than beginner ($M = 3.62$, $SE = 0.11$) and expert-level notifications (($M = 3.81$, $SE = 0.10$), $F(1, 129) = 48.47$, $p < .001$, partial $\eta^2 = 0.27$. Again, this pattern was consistently observed in participants with both high and low levels of technical proficiency.

Although no significant main effect of notification length was found, a small interaction effect between notification complexity and length emerged, $F(1, 129) = 23.66$, $p < .001$, partial $\eta^2 = 0.16$. The interaction was constituted by the pattern that at the beginner level, longer notifications ($M = 3.82$, $SE = 0.12$) elicited higher motivation to act than shorter notifications ($M = 3.41$, $SE = 0.12$), $t(129) = 4.88$, $p < .001$, $d_z = 0.43$. In contrast, at the expert level, shorter notifications ($M = 3.87$, $SE = 0.11$) elicited marginally higher motivation to act than longer notifications ($M = 3.76$, $SE = 0.12$), however, the difference was not statistically significant, $t(129) = 1.48$, $p = .071$, $d_z = 0.13$.
	

\begin{figure}[h]
    \centering
    \includegraphics[width=.7\textwidth]{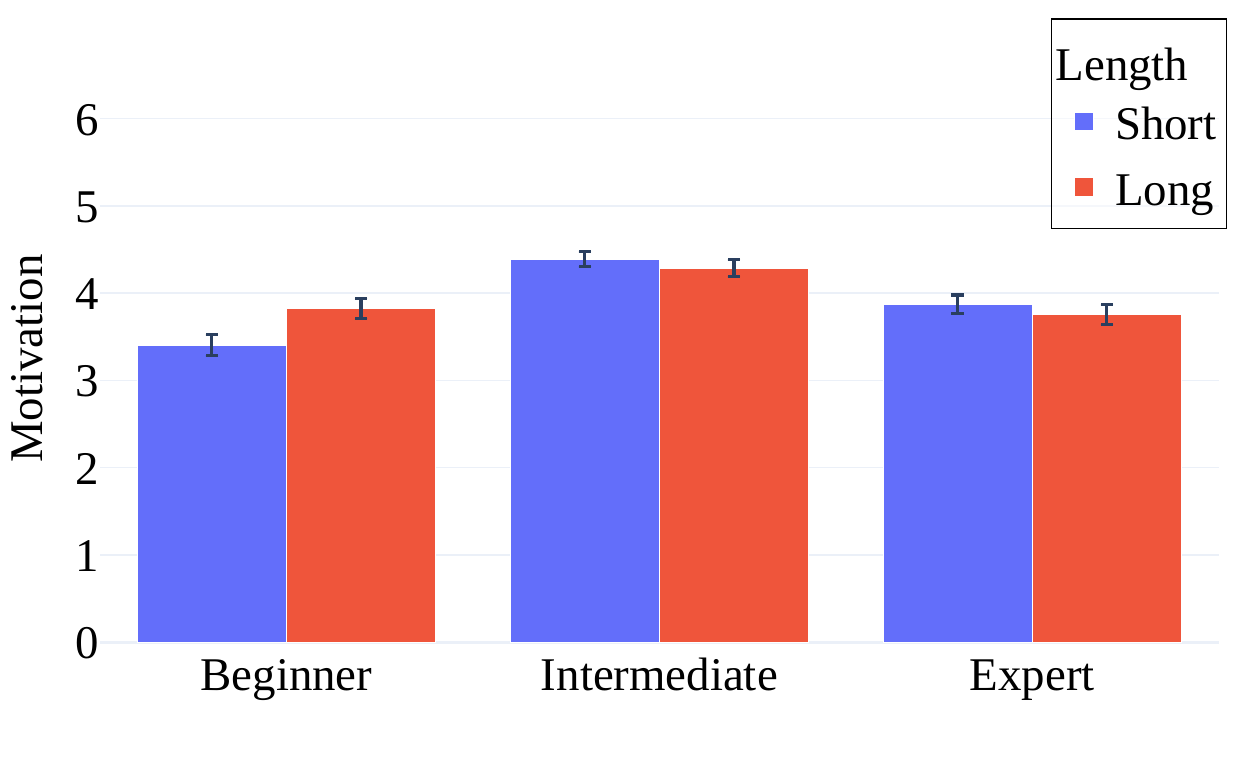}
    \caption{Motivation-to-act ratings by notification complexity and length, showing intermediate notifications as most motivating. Error bars represent standard errors of the mean.}
    \label{fig:motivation}
\end{figure}

In summary, our analysis showed that intermediate-level notifications generated the highest motivation to act, followed by expert and beginner-level notifications. Notification length did not have a significant overall effect. We observed a significant interaction (see Figure~\ref{fig:motivation}) at beginner level. Longer notifications motivated participants more. At the expert level, shorter notifications gave participants a slight edge, although this difference was not statistically significant.

\section{Discussion}
\label{sec:discussion}



\subsection{Interpreting Our Findings}

Our findings indicate that intermediate-level notifications were the most effective, showing higher ratings for likability, understandability, and motivation to act compared to beginner or expert-level notifications. This consistent preference for intermediate-level notifications cannot be attributed to participants' level of technical proficiency, as both those with high and low levels of proficiency consistently rated intermediate-level notifications as the most favorable.

The influence of length varied by notification complexity. Longer notifications were perceived as more likeable and elicited higher motivation to act at the beginner-level, whereas shorter notifications were perceived as marginally more likeable, more understandable, and elicited marginally higher motivation to act at the expert-level.
Our findings provide 
evidence for how to communicate cybersecurity risks in ways that are both accessible and engaging. 
The fact that intermediate complexity performed well across user groups suggests that it may serve as a 
default in settings where 
expertise is unknown. 

\subsection{Design and Practical Implications}

\paragraph{Design Implications}

\textbf{D1: Use intermediate complexity as the default.}  
Notifications written at an intermediate level of complexity were rated most effective across all user groups. When a system cannot reliably assess user expertise, defaulting to intermediate complexity offers a robust default setting.

\textbf{D2: Support adaptive messaging based on user proficiency.}  
When users are willing to disclose their technical background, systems can personalize notifications accordingly. Expert users may benefit from concise, high-signal messages, while beginner users may need additional context and reassurance. 

\paragraph{Practical Implications}

\textbf{P1: Replace raw alerts with templated, LLM-generated messages.}  
Current smart home IDS solutions often surface raw technical alerts, which are difficult for non-experts to interpret. Our results support the integration of LLMs and structured templates to transform these alerts into messages that users can understand and act upon, increasing the likelihood of timely responses.

\textbf{P2: Include minimal structured metadata in alerts.}  
Providing just two pieces of structured information—device type and attack type—enables LLMs to generate meaningful and actionable notifications. IDS developers should prioritize making this metadata available at runtime, especially for consumer systems.

\subsection{Limitations and Future Work}


\paragraph{L1: Ecological validity.}  
Our experimental user study was conducted in a controlled, online setting using hypothetical scenarios. While this allowed us to isolate specific design variables, it may not reflect how users respond to security notifications in real-world smart home contexts—especially under time pressure, stress, or perceived risk. Future work should explore how users engage with notifications during actual security events, and whether guidance is successfully followed and leads to meaningful outcomes.

\paragraph{L2: Participant sample bias.}  
Our Prolific sample skewed toward technically proficient participants, likely due to the study title (“Cybersecurity Notifications”) and self-selection effects. Notably, no participant reported “very low” technical proficiency. The sample also lacked broader demographic diversity, with a majority from the UK and an overrepresentation of male participants. Future studies should aim to recruit more diverse populations, including those with lower technical skill levels and varied cultural or linguistic backgrounds.

\paragraph{L3: Variation in LLM outputs.}  
All security notifications in our study were generated using GPT-4o, which ensured consistency but may limit generalizability to other LLMs. Future work should evaluate whether similar effects hold when using alternative models, including smaller or locally hosted systems (e.g., Phi, Gemma), which may be more suitable for privacy-sensitive environments.

\paragraph{L4: Messages at the beginner level.}  
Some beginner-level notifications included awkward or overly generic phrases (e.g., “consult a professional cybersecurity expert”). While we preserved these to maintain internal consistency, they may have affected ratings. Future studies should iteratively refine beginner-level phrasing using user-centered techniques such as think-aloud protocols or co-design.

\paragraph{L5: Alert quality and false positives.}  
Intrusion Detection Systems, especially those using anomaly detection, often produce a high volume of false positives. While our study assumed reasonably accurate alerts, future work should explore how notification design can mitigate alert fatigue and help users build  trust in the system—especially when alerts are noisy, ambiguous, or misleading.

\paragraph{Future Directions.}  
Taken together, these limitations highlight the need for more ecologically valid, longitudinal research that examines how actionable notifications perform in diverse, real-world settings. Future work should investigate how personalization, timing, and repetition influence user behavior over time, and how to design notifications that balance clarity, urgency, and user autonomy without causing fatigue. Finally, exploring how LLM-generated notifications affect user trust and security outcomes—especially when models make errors—remains an essential open challenge.

\section{Conclusion}
\label{sec:conclusion}

Our study highlights the importance of tailoring actinable LLM-generated security notifications to smart home users with varying levels of technical proficiency. By conducting an experimental study with 130 participants, we found that intermediate-complexity notifications are the most effective across all user groups, consistently receiving the highest ratings in likability, understandability, and motivation to act. The effect of length varied by notification type: longer versions of beginner-level messages were more effective, while shorter versions of expert-level messages were marginally preferred.  These findings provide insights for designing actionable security notifications that balance clarity and engagement, ensuring that smart home users can effectively respond to potential cyber threats. Future work should explore real-world deployment scenarios, investigate adaptive notification mechanisms, and assess the effectiveness of different LLM models to further refine security notification design.
\paragraph{Acknowledgments}

The authors acknowledge financial support from the Federal Ministry of Education and Research of Germany and from the Sächsische Staatsministerium für Wissenschaft, Kultur und Tourismus within the program Center of Excellence for AI Research: ``Center for Scalable Data Analytics and Artificial Intelligence Dresden/Leipzig,'' project identification number: ScaDS.AI.

\bibliographystyle{template/bibtex/splncs03_unsrt}
\bibliography{literature}

\begin{thebibliography}{10}
\providecommand{\url}[1]{\texttt{#1}}
\providecommand{\urlprefix}{URL }

\bibitem{BUILGIL2023107770}
Buil-Gil, D., Kemp, S., Kuenzel, S., Coventry, L., Zakhary, S., Tilley, D., Nicholson, J.: The digital harms of smart home devices: A systematic literature review. Computers in Human Behavior  145,  107770 (2023)

\bibitem{AnthiIDS2018}
Anthi, E., Williams, L., Słowińska, M., Theodorakopoulos, G., Burnap, P.: A supervised intrusion detection system for smart home iot devices. IEEE Internet of Things Journal  6(5),  9042--9053 (2019)

\bibitem{snort_ruleset}
{Cisco Systems}: Snort community ruleset. \url{https://www.snort.org/downloads/\#rule-downloads} (2025), accessed: 2025-04-08

\bibitem{suricata_ruleset}
{Open Information Security Foundation}: Suricata ruleset. \url{https://github.com/OISF/suricata/tree/master/rules} (2025), accessed: 2025-04-08

\bibitem{et_ruleset}
{Proofpoint, Inc.}: Emerging threats rules. \url{https://rules.emergingthreats.net/} (2025), accessed: 2025-04-08

\bibitem{juettner2024chatids}
J\"{u}ttner, V., Grimmer, M., Buchmann, E.: Chatids: Advancing explainable cybersecurity using generative ai. Journal On Advances in Security  17(1,2) (2024)

\bibitem{hoffmann2024ChatSEC}
Hoffmann, M., Buchmann, E.: Chatsec: Spicing up vulnerability scans with ai for heterogeneous university it - towards enhancing security vulnerability reports for non-experts. In: Proceedings of the 1st International Conference on AI-based Systems and Services (AISyS'24) (2024)

\bibitem{Boss2015}
Boss, S., Galletta, D., Lowry, P.B., Moody, G.D., Polak, P.: What do systems users have to fear? using fear appeals to engender threats and fear that motivate protective security behaviors. MIS Quarterly (MISQ)  39(4),  837--864 (2015)

\bibitem{openai2024gpt4o}
{OpenAI}: Hello gpt-4o. \url{https://openai.com/index/hello-gpt-4o/} (2024), accessed: 2025-01-25

\bibitem{deLeeuw2023}
de~Leeuw, J.R., Gilbert, R.A., Luchterhandt, B.: jspsych: Enabling an open-source collaborative ecosystem of behavioral experiments. Journal of Open Source Software  8(85),  5351 (2023), \url{https://doi.org/10.21105/joss.05351}

\bibitem{lange2015jatos}
Lange, K., Kühn, S., Filevich, E.: “just another tool for online studies” (jatos): An easy solution for setup and management of web servers supporting online studies. PLoS ONE  10(6),  e0130834 (2015), \url{https://doi.org/10.1371/journal.pone.0130834}

\bibitem{mindprobe}
LindenLoot, CogSci: Mindprobe. \url{https://mindprobe.eu/} (2024), accessed: 2025-01-25

\bibitem{smart-device-vulns}
Chhetri, C., Motti, V.: Identifying vulnerabilities in security and privacy of smart home devices. In: Choo, K.K.R., Morris, T., Peterson, G.L., Imsand, E. (eds.) National Cyber Summit (NCS) Research Track 2020. pp. 211--231. Springer International Publishing, Cham (2021)

\bibitem{smart-network-issues}
Touqeer, H., Zaman, S., Amin, R., Hussain, M., Al-Turjman, F., Bilal, M.: Smart home security: challenges, issues and solutions at different iot layers. J. Supercomput.  77(12),  14053–14089 (dec 2021)

\bibitem{Pattnaik2022ASO}
Pattnaik, N., Li, S., Nurse, J.R.C.: A survey of user perspectives on security and privacy in a home networking environment. ACM Computing Surveys  55,  1 -- 38 (2022)

\bibitem{nist-ids}
Scarfone, K., Mell, P.: Guide to intrusion detection and prevention systems (idps). \url{https://tsapps.nist.gov/publication/get_pdf.cfm?pub_id=50951} (2007)

\bibitem{cranor-human-in-loop}
Cranor, L.F.: A framework for reasoning about the human in the loop. In: Proceedings of the 1st Conference on Usability, Psychology, and Security. UPSEC'08, USENIX Association, USA (2008)

\bibitem{ZIMMERMANN2019169}
Zimmermann, V., Renaud, K.: Moving from a ‘human-as-problem” to a ‘human-as-solution” cybersecurity mindset. International Journal of Human-Computer Studies  131,  169--187 (2019)

\bibitem{rogers1983cognitive}
Rogers, R.W.: Cognitive and physiological processes in fear appeals and attitude change: A revised theory of protection motivation. Social psychology: A source book pp. 153--176 (1983)

\bibitem{wogalter2018communication}
Wogalter, M.S.: Communication-human information processing (c-hip) model. In: Forensic human factors and ergonomics, pp. 33--49. CRC Press (2018)

\bibitem{warning-study}
Zaaba, Z.F., Lim Xin~Yi, C., Amran, A., Omar, M.A.: Harnessing the challenges and solutions to improve security warnings: A review. Sensors  21(21) (2021)

\bibitem{scaring-warning}
Sasse, A.: Scaring and bullying people into security won't work. IEEE Security and Privacy  13(3),  80–83 (2015)

\bibitem{warning-eeg}
Conrad, C., Aziz, J., Smith, N., Newman, A.: What do users feel? towards affective eeg correlates of cybersecurity notifications. In: Davis, F.D., Riedl, R., vom Brocke, J., L{\'e}ger, P.M., Randolph, A.B., Fischer, T. (eds.) Information Systems and Neuroscience. pp. 153--162. Springer International Publishing, Cham (2020)

\bibitem{zimmermann2024beyondFear}
Von~Preuschen, A., Schuhmacher, M.C., Zimmermann, V.: Beyond fear and frustration - towards a holistic understanding of emotions in cybersecurity. In: Proceedings of the Twentieth USENIX Conference on Usable Privacy and Security. SOUPS '24, USENIX Association, USA (2024)

\bibitem{microsoft2024copilot}
Microsoft: Microsoft copilot for security. \url{https://learn.microsoft.com/en-us/copilot/security/} (2024), accessed: 2025-01-25

\bibitem{YAO2024100211}
Yao, Y., Duan, J., Xu, K., Cai, Y., Sun, Z., Zhang, Y.: A survey on large language model (llm) security and privacy: The good, the bad, and the ugly. High-Confidence Computing  4(2),  100211 (2024)

\bibitem{stochastic-parrots}
Bender, E.M., Gebru, T., McMillan-Major, A., Shmitchell, S.: On the dangers of stochastic parrots: Can language models be too big? In: Proceedings of the 2021 ACM Conference on Fairness, Accountability, and Transparency. p. 610–623. FAccT '21, Association for Computing Machinery, New York, NY, USA (2021)

\bibitem{Hicks2024}
Hicks, M.T., Humphries, J., Slater, J.: Chatgpt is bullshit. Ethics and Information Technology  26(2), ~38 (2024)

\bibitem{aegis}
Sikder, A.K., Babun, L., Uluagac, A.S.: Aegis+: A context-aware platform-independent security framework for smart home systems. Digital Threats  2(1) (2021)

\bibitem{draf}
Collen, A., Nijdam, N.A.: Can i sleep safely in my smarthome? a novel framework on automating dynamic risk assessment in iot environments. Electronics  11(7) (2022)

\bibitem{piti}
Visoottiviseth, V., Chutaporn, G., Kungvanruttana, S., Paisarnduangjan, J.: Piti: Protecting internet of things via intrusion detection system on raspberry pi. In: 2020 International Conference on Information and Communication Technology Convergence (ICTC). pp. 75--80 (2020)

\bibitem{user-centric-threat-model}
Datta, P., Sartoli, S., Gutierrez, L.F., Abri, F., Namin, A.S., Jones, K.S.: A user-centric threat model and repository for cyber attacks. In: Proceedings of the 37th ACM/SIGAPP Symposium on Applied Computing. p. 1341–1346. SAC '22, Association for Computing Machinery, New York, NY, USA (2022)

\bibitem{ghost}
Collen, A., Nijdam, N.A., Augusto-Gonzalez, J., Katsikas, S.K., Giannoutakis, K.M., Spathoulas, G., Gelenbe, E., Votis, K., Tzovaras, D., Ghavami, N., Volkamer, M., Haller, P., S{\'a}nchez, A., Dimas, M.: Ghost - safe-guarding home iot environments with personalised real-time risk control. In: Gelenbe, E., Campegiani, P., Czach{\'o}rski, T., Katsikas, S.K., Komnios, I., Romano, L., Tzovaras, D. (eds.) Security in Computer and Information Sciences. pp. 68--78. Springer International Publishing, Cham (2018)

\bibitem{internet-of-threats-ghost}
Augusto-Gonzalez, J., Collen, A., Evangelatos, S., Anagnostopoulos, M., Spathoulas, G., Giannoutakis, K.M., Votis, K., Tzovaras, D., Genge, B., Gelenbe, E., Nijdam, N.A.: From internet of threats to internet of things: A cyber security architecture for smart homes. In: 2019 IEEE 24th International Workshop on Computer Aided Modeling and Design of Communication Links and Networks (CAMAD). pp. 1--6 (2019)

\bibitem{hub2027}
Simpson, A.K., Roesner, F., Kohno, T.: Securing vulnerable home iot devices with an in-hub security manager. In: 2017 IEEE International Conference on Pervasive Computing and Communications Workshops (PerCom Workshops). pp. 551--556 (2017)

\bibitem{iotfla}
Aïvodji, U.M., Gambs, S., Martin, A.: Iotfla : A secured and privacy-preserving smart home architecture implementing federated learning. In: 2019 IEEE Security and Privacy Workshops (SPW). pp. 175--180 (2019)

\bibitem{snort_classtypes}
{Cisco Systems}: Snort classtypes. \url{https://docs.snort.org/rules/options/general/classtype} (2024), accessed: 2025-01-25

\bibitem{faul2009statistical}
Faul, F., Erdfelder, E., Buchner, A., Lang, A.G.: Statistical power analyses using g*power 3.1: Tests for correlation and regression analyses. Behavior Research Methods  41(4),  1149--1160 (2009), \url{https://doi.org/10.3758/BRM.41.4.1149}

\bibitem{cohen1988}
Cohen, J.: Statistical Power Analysis for the Behavioral Sciences. Lawrence Erlbaum Associates, Inc., Hillsdale, NJ, 2nd edn. (1988)

\bibitem{echo-dot}
{Amazon.com, Inc.}: Echo dot (5th gen, 2022 release). \url{https://www.amazon.com} (2024), accessed: 2025-01-25

\bibitem{ring-doorbell}
{Ring LLC}: Ring video doorbell (2nd generation). \url{https://ring.com} (2024), accessed: 2025-01-25

\bibitem{schlage-lock}
{Allegion plc}: Schlage encode smart wifi deadbolt. \url{https://www.schlage.com} (2024), accessed: 2025-01-25

\bibitem{blink-indoor}
{Immedia Semiconductor LLC}: Blink indoor (3rd gen). \url{https://blinkforhome.com} (2024), accessed: 2025-01-25

\bibitem{geeni-sensor}
{Merkury Innovations}: Geeni door/window sensor. \url{https://mygeeni.com} (2024), accessed: 2025-01-25

\bibitem{kasa-bulb}
{TP-Link}: Kasa smart light bulb. \url{https://www.kasasmart.com} (2024), accessed: 2025-01-25

\bibitem{wyze-plug}
{Wyze Labs, Inc.}: Wyze plug. \url{https://www.wyze.com} (2024), accessed: 2025-01-25

\bibitem{bosch-thermostat}
{Robert Bosch GmbH}: Smart radiator thermostat ii. \url{https://www.bosch-smarthome.com} (2024), accessed: 2025-01-25

\bibitem{shelly-ht}
{Allterco Robotics}: Shelly ht gen3. \url{https://www.shelly.com} (2024), accessed: 2025-01-25

\bibitem{shark-matrix}
{SharkNinja Operating LLC}: Shark matrix. \url{https://www.sharkclean.com} (2024), accessed: 2025-01-25

\end{thebibliography}

\appendix

\section{Security Notifications}
\label{sec:sec-notifications}

\begin{center}
\begin{tcolorbox}[breakable, enhanced, sharp corners, colback=white, colframe=black, boxrule=0.5pt, left=1mm, right=1mm, top=1mm, bottom=1mm]
\small
Hello,\\
Your smart home security system has found a possible security problem with your Amazon Echo Dot 5th Gen, a smart speaker. Please follow these suggestions to fix the issue and secure your device and network:

\begin{itemize}
    \item \textbf{Turn off the Echo Dot}: Unplug your Amazon Echo Dot 5th Gen from the power outlet to stop the network scan that has been detected.
    \item \textbf{Contact a Professional}: Reach out to a trusted friend or a professional technician for further assistance to investigate and resolve the intrusion issue.
\end{itemize}

Best regards,\\
Your Smart Home Security Team
\end{tcolorbox}
\captionof{figure}{Short Beginner Security Notification}
\end{center}

\begin{center}
\begin{tcolorbox}[breakable, enhanced, sharp corners, colback=white, colframe=black, boxrule=0.5pt, left=1mm, right=1mm, top=1mm, bottom=1mm]
\small
Hello,\\
Your smart home security system has found a possible security problem with your Amazon Echo Dot 5th Gen, a smart voice-controlled assistant. Please follow these suggestions to fix the issue and secure your device and network:

\begin{itemize}
    \item \textbf{Disconnect Device}: Unplug the Amazon Echo Dot 5th Gen from the power source to stop any potential malicious activity immediately.
    \item \textbf{Reset Device}: Perform a factory reset on the Amazon Echo Dot 5th Gen via the Alexa app or by using the device's hardware reset button.
    \item \textbf{Change Passwords}: Change the passwords for your Amazon account and Wi-Fi network to new, strong, and unique passwords to prevent unauthorized access.
    \item \textbf{Update Software}: Reconnect and ensure that the Amazon Echo Dot 5th Gen has the latest firmware updates installed via the Alexa app.
\end{itemize}

Best regards,\\
Your Smart Home Security Team
\end{tcolorbox}
\captionof{figure}{Long Intermediate Security Notification}
\end{center}


\begin{center}
\begin{tcolorbox}[breakable, enhanced, sharp corners, colback=white, colframe=black, boxrule=0.5pt, left=1mm, right=1mm, top=1mm, bottom=1mm]
\small
Hello,\\
Your smart home security system has found a possible security problem with your Amazon Echo Dot 5th Gen, a smart speaker. Please follow these suggestions to fix the issue and secure your device and network:

\begin{itemize}
    \item \textbf{Inspect Network Traffic and Device}: Verify device traffic using network monitoring tools to identify any irregularities, and reboot the Amazon Echo Dot 5th Gen to reset its network state.
    \item \textbf{Update Firmware and Strengthen Security}: Ensure the Amazon Echo Dot 5th Gen firmware is updated. Check for router firmware updates. Strengthen network security by updating administrator passwords and configuring a WPA3 encryption protocol.
\end{itemize}

Best regards,\\
Your Smart Home Security Team
\end{tcolorbox}
\captionof{figure}{Short Expert Security Notification}
\label{fig:sec-notification-advanced}
\end{center}

\section{Prompt Templates}
\label{sec:prompt}

\begin{center}
\begin{tcolorbox}[breakable, enhanced, sharp corners, colback=white, colframe=black, boxrule=0.5pt, left=1mm, right=1mm, top=1mm, bottom=1mm]
\small
User Jon has a smart home equipped with a cybersecurity system. The system monitors the network traffic of all smart home devices in Jon's smart home and sends an alert when it detects an intrusion. Right now, the intrusion [ALERT] has been detected on the device [DEVICE]. Your task is to assume the role of a cybersecurity expert and provide useful instructions on how Jon can fix the problem on the affected device and on his network. Choose instructions based on the threat level of the alert. Select instructions according to Jon's skill level. Jon is not technologically savvy; he can perform simple tasks like turning the affected device off, turning the router off and on, or asking for help. Jon can't change passwords, install updates or check connections. Provide simple instructions on how Jon can fix the problem on his device and secure his network. The instructions should have only two steps: one step is your recommended instruction and one step that suggests getting help. Each step in the instruction must have 30 words and titles instead of numbers. The message you generate must have the same structure as this example. Replace the device name and short device description, and create the instructions as an unordered list:\\  
Hello,

Your smart home security system has found a possible security problem with your Ring Video Doorbell 2nd Gen, a smart video doorbell. Please follow these suggestions to fix the issue and secure your device and network:

- **Restart the Device**: Turn off the Ring Video Doorbell 2nd Gen by unplugging it or removing the battery. Wait one minute and turn it back on to reset connections.  

- **Seek Assistance**: If the issue continues, ask a tech-savvy friend or a professional to check your device and network for possible security threats and provide the necessary help.  

Best regards,\\  
Your Smart Home Security Team
\end{tcolorbox}
\captionof{figure}{Prompt template for a short-beginner security notification}
\label{fig:basic-security-template}
\end{center}


\begin{center}
\begin{tcolorbox}[breakable, enhanced, sharp corners, colback=white, colframe=black, boxrule=0.5pt, left=1mm, right=1mm, top=1mm, bottom=1mm]
\small
User Jon has a smart home equipped with a cybersecurity system. The system monitors the network traffic of all smart home devices in Jon's smart home and sends an alert when it detects an intrusion. Right now, the intrusion [ALERT] has been detected on the device [DEVICE]. Your task is to assume the role of a cybersecurity expert and provide useful instructions on how Jon can fix the problem on the affected device and on his network. Choose instructions based on the threat level of the alert. Select instructions according to Jon's skill level. Jon has a basic understanding of the smart home and can perform tasks such as resetting the router, resetting devices, changing passwords and installing updates. Jon doesn't know how to disable protocols, change encryption or do network scans. The instructions should have no more than two steps. Each step in the instruction must have 30 words and titles instead of numbers. The message you generate must have the same structure as this example. Replace the device name and short device description, and create the instructions as an unordered list:\\ 
Hello,

Your smart home security system has found a possible security problem with your Ring Video Doorbell 2nd Gen, a smart video doorbell. Please follow these suggestions to fix the issue and secure your device and network:

- **Title**: Explanation

Best regards,\\
Your Smart Home Security Team
\end{tcolorbox}
\captionof{figure}{Prompt template for a short-intermediate security notification}
\label{fig:intermediate-short-template}
\end{center}


\begin{center}
\begin{tcolorbox}[breakable, enhanced, sharp corners, colback=white, colframe=black, boxrule=0.5pt, left=1mm, right=1mm, top=1mm, bottom=1mm]
\small
User Jon has a smart home equipped with a cybersecurity system. The system monitors the network traffic of all smart home devices in Jon's smart home and sends an alert when it detects an intrusion. Right now, the intrusion [ALERT] has been detected on the device [DEVICE]. Your task is to assume the role of a cybersecurity expert and provide useful instructions on how Jon can fix the problem on the affected device and on his network. Choose instructions based on the threat level of the alert. Select instructions according to Jon's skill level. Jon works in IT and has a broad knowledge of technology. Provide instructions on how Jon can resolve the issue on his device and secure his network. The instructions should have no more than two steps. Each step in the instruction must have 30 words and titles instead of numbers. The message you generate must have the same structure as this example. Replace the device name and short device description, and create the instructions as an unordered list:\\  
Hello,

Your smart home security system has found a possible security problem with your Ring Video Doorbell 2nd Gen, a smart video doorbell. Please follow these suggestions to fix the issue and secure your device and network:

- **Analyze Network Logs**: Check router logs and firewall settings to identify suspicious activity from the Ring Video Doorbell. Look for unusual connections, blocked requests, or repeated authentication failures.  

- **Update and Secure**: Update the Ring Video Doorbell firmware, change device credentials, and enable two-factor authentication if available. Configure router settings to restrict device access and limit unnecessary external communication.  

Best regards,\\  
Your Smart Home Security Team
\end{tcolorbox}
\captionof{figure}{Prompt template for an short-expert  security notification}
\label{fig:advanced-security-template}
\end{center}

\section{Snort3-Community Alert Messages}
\label{sec:alert-messages}

\newcolumntype{L}[1]{>{\raggedright\arraybackslash}p{#1}}

\begin{longtable}{@{}p{1.5cm}L{4cm}p{5cm}c@{}}
\toprule
\textbf{SID} & \textbf{Alert Message} & \textbf{Explanation} & \textbf{Severity} \\
\midrule
\endfirsthead

\toprule
\textbf{SID} & \textbf{Alert Message} & \textbf{Explanation} & \textbf{Severity} \\
\midrule
\endhead

\midrule
\multicolumn{4}{r}{\textit{Continued on next page}} \\
\midrule
\endfoot

\bottomrule
\caption{Alert Messages from the Snort3-Community Ruleset} \label{tab:alert-messages} \\
\endlastfoot

1:484 & PROTOCOL-ICMP PING Sniffer Pro/NetXRay network scan & Indicates that a tool is scanning the network by sending ping requests to various devices. This can be done to map devices on the network but can also signify probing. & 3 \\
1:26563 & MALWARE-CNC Harakit botnet traffic & Indicates traffic from a device that might be part of the Harakit botnet. & 1 \\
1:1638 & INDICATOR-SCAN SSH Version map attempt & Shows an attempt to identify the version of SSH running on a device. This is often used to find vulnerabilities that can be exploited. & 3 \\
1:33648 & MALWARE-CNC Linux.Trojan.XORDDoS outbound connection & Indicates that a device is infected by the XORDDoS Trojan, which is a type of malware that infects Linux systems and is often used to launch DDoS attacks. & 1 \\
1:402 & PROTOCOL-ICMP destination unreachable port unreachable packet detected & Indicates that a device is informing the sender that a specific port is unreachable. It could be a normal network communication or a sign of probing. & 3 \\
1:3147 & PROTOCOL-TELNET login buffer overflow attempt & Shows an attempt to exploit a vulnerability in the Telnet service to gain unauthorized access. Such attempt could crash the service or allow control over the device. & 1 \\
1:467 & PROTOCOL-ICMP Nemesis v1.1 Echo & Shows that the tool Nemesis, is sending echo requests to devices. It might be part of network testing or scanning activities. & 2 \\
1:215 & MALWARE-BACKDOOR MISC Linux rootkit attempt & Indicates an attempt to install a rootkit on a Linux device. Rootkits can hide malicious activities and give attackers persistent access to the device. & 1 \\
1:492 & PROTOCOL-TELNET login failed & This alert is triggered when a login attempt to a Telnet service fails. Multiple failed attempts could indicate someone trying to guess passwords. & 2 \\
1:210 & MALWARE-BACKDOOR attempt & Shows a general attempt to exploit a backdoor, a method to bypass normal authentication. & 1 \\

\end{longtable}

\section{Smart Home Devices}
\label{sec:devices}

\begin{table}[h]
\centering
\begin{tabular}{llp{7cm}}
\toprule
\textbf{} & \textbf{Device} & \textbf{Description} \\
\midrule
A & Amazon Echo Dot 5th Gen~\cite{echo-dot} & Smart speaker with microphone and speaker for various functionalities. \\
B & Ring Video Doorbell 2nd Gen~\cite{ring-doorbell} & Smart doorbell with camera, microphone, and speaker. \\
C & Schlage Encode Smart Lock~\cite{schlage-lock} & Smart lock with remote access. \\
D & Blink Indoor 3rd Gen~\cite{blink-indoor} & Smart security camera with motion detection and microphone. \\
E & Geeni Door/Window Sensor~\cite{geeni-sensor} & Smart sensor detecting door/window openings and sending alerts. \\
F & Kasa Smart Light Bulb~\cite{kasa-bulb} & Smart light bulb with dimming and scheduling features. \\
G & Wyze Plug~\cite{wyze-plug} & Smart plug for remote control of devices. \\
H & Smart radiator thermostat II~\cite{bosch-thermostat} & Smart thermostat for radiators with remote temperature control. \\
I & Shelly H\&T Gen3~\cite{shelly-ht} & Smart sensor monitoring humidity and temperature. \\
J & Shark Matrix~\cite{shark-matrix} & Smart robot vacuum with precise navigation. \\
\bottomrule
\end{tabular}
\caption{Smart Devices used in the Study. All devices connect over Wi-Fi and can be controlled via smartphone.}
\label{tab:all-devices}
\end{table}

\end{document}